\documentclass[
twocolumn,
]{ceurart}

\usepackage{listings}
\begin{document}

\copyrightyear{2024}
\copyrightclause{Copyright © 2024 for this paper by its authors. Use permitted under Creative Commons License Attribution 4.0 International (CC BY 4.0).}

\conference{Published in the Workshop Proceedings of the EDBT/ICDT 2024 Joint Conference (March 25-28, 2024, Paestum, Italy).}

\title{Dataversifying Natural Sciences: \\Pioneering a Data Lake Architecture for Curated Data-Centric Experiments in Life \& Earth Sciences}



\author[1]{Genoveva Vargas-Solar}[%
email=genoveva.vargas-solar@cnrs.fr,
]
\address[1]{CNRS, Univ. Lyon,  INSA Lyon, UCBL, LIRIS, UMR5205, F-69221, France}

\author[2]{Jérôme Darmont}[%
email=jerome.darmont@univ-lyon2.fr,
]

\author[4]{Alejandro Adorjan}[%
email=aadorian@gmail.com,
]

\author[1,3]{ Javier A. Espinosa-Oviedo}[%
email=javier.espinosa@liris.cnrs.fr,
]

\author[5]{Carmem Hara}[%
email=carmemhara@ufpr.br,
]

\author[2]{Sabine Loudcher}[%
email=sabine.loudcher@univ-lyon2.fr,
]

\author[6]{Regina Motz}[%
email=rmotz@fing.edu.uy,
]

\author[7]{Martin Musicante}[%
email=mam@dimap.ufrn.br,
]

\author[8]{José-Luis Zechinelli-Martini}[%
email=joseluis.zechinelli@udlap.mx,
]

\address[2]{Université de Lyon, Lyon 2, UR ERIC  5 avenue Mendès France, 69676 Bron Cedex, France
 }

\address[3]{CPE Lyon, 43 Blvd. du 11 Novembre 1918, 69616 Villeurbanne Cedex, France}

\address[4]{Unversidad ORT, Montevideo, Uruguay }

\address[5]{Universidade Federal do Paranà, Dept. de Informatica, Curitiba - PR, 81531-980, Brazil }

\address[6]{Instituto de Computación (INCO) Facultad de Ingeniería, Universidad de la Repúbica, Uruguay}

\address[7]{Universidad Federal Rio Grande do Norte, DIMAP, Natal, Brazil }

\address[8]{Fundación Universidad de las Américas, Puebla
Exhacienda Sta. Catarina Mártir s/n  72820 San Andrés Cholula, Mexico }

\cortext[1]{Genoveva Vargas-Solar.}
\fntext[1]{The authors' list is alphabetical except for the first two authors.}

\begin{abstract}
This vision paper introduces a pioneering data lake architecture designed to meet Life \& Earth sciences' burgeoning data management needs. As the data landscape evolves, the imperative to navigate and maximise scientific opportunities has never been greater. Our vision paper outlines a strategic approach to unify and integrate diverse datasets, aiming to cultivate a collaborative space conducive to scientific discovery.
The core of the design and construction of a data lake is the development of formal and semi-automatic tools, enabling the meticulous curation of quantitative and qualitative data from experiments. Our unique "research-in-the-loop" methodology ensures that scientists across various disciplines are integrally involved in the curation process, combining automated, mathematical, and manual tasks to address complex problems, from seismic detection to biodiversity studies. By fostering reproducibility and applicability of research, our approach enhances the integrity and impact of scientific experiments. This initiative is set to improve data management practices, strengthening the capacity of Life \& Earth sciences to solve some of our time's most critical environmental and biological challenges.
\end{abstract}


\maketitle

\section{Introduction}


These days, it is relatively easy and inexpensive to acquire massive amount of data, even in continuous mode. This has been no different for experimental and observational sciences like Life \& Earth sciences.
%
%
Accessibility to data
about the Earth and its biodiversity, with varying levels of provenance, quality and reliability, opens up the possibility of constructing different perspectives on the phenomena observed, leading to scientific conclusions with different depths that target a wide range of knowledge consumers (civilians, decision-makers, scientists). 

Traditional \emph{schema-on-write} approaches, such as the Extraction, Transformation and Loading (ETL) process, are ineffective for the data management requirements of these
experimental sciences. Data lakes are becoming increasingly common for the management and analysis of massive data. Data lakes are repositories that store raw data in its original format. They can be well adapted for storing data harvested from digital sources (observation stations), social media, Web and in situ collectors.

The extraction of value through data-driven experiments in the Life \& Earth sciences is
determined by two main elements:

\begin{itemize}

    \item The maintenance of metadata gathering the conditions under which experiments are performed (quantitative perspective) to preserve the memory of the experimental process of knowledge production process, and to enable understanding and reproducibility.

    \item An open science perspective that can go beyond data sharing and must consider the sharing of know-how, decision-making, expertise, project management, and people within the projects that define the research must be considered.
    
\end{itemize}

This vision paper introduces our approach to designing and building a data lake for collecting and integrating data and meta data of Life \& Earth sciences' data-driven experiments. 

The remainder of the paper is organised as follows. Section \ref{sec:relatedwork} gives a general overview of approaches that address curating and managing knowledge in Life \& Earth sciences. 
Section \ref{sec:approach} describes the challenges associated with curating data and data-driven experiments in Life \& Earth sciences often guided by researchers. In particular, the section gives the general challenges for building data lakes containing curated data and producing knowledge derived from data-driven experiments. 
Section \ref{sec:platform} introduces the general principle for building, maintaining and exploiting a data lake. The data lake allows the creation of "dataverses" that can export the history of the development of experimental processes that lead to knowledge in Life \& Earth sciences. Finally, Section \ref{sec:conclusion} concludes the paper and discusses future work.

\section{Related work}\label{sec:relatedwork}

We introduce the main topics and approaches that underline the vision of maintaining and sharing data to perform data-driven experiments: data harvesting tools, data curation techniques, data labs, data lakes, science lakes and dataverses.

\subsection{Data harvesting}
Data available on the Web play a determining role in decision-making in personal and corporate life. Collecting and storing this data in a structured model helps integrate them with other sources and use the dataset in various applications, such as event detection and sentiment monitoring. Online newspapers are essential sources of information, accessed daily by thousands of people. 

Various works in the literature report manual efforts to extract data from pages on the \textit{Web} \cite{vargas2021laclichev, nascimento2022redes}. However, these efforts have been eased by applying Web scraping techniques. 
Some work complements automated extraction processes to obtain clean and analysed data by implementing curation procedures \cite{sarr2018factextract}.
Among the various existing tools available on the \textit{Web} for data extraction, we can highlight ParseHub\footnote{\url{https://www.parsehub.com/}} is a web scraping tool that facilitates data extraction from websites through an interactive click-based interface, saving the data directly to the cloud in JSON and CSV formats. It navigates through continuation pages and captures complete news articles, with the ability to collect data based on specific character sequences. 80legs\footnote{\url{https://80legs.com/} } offers sequential data extraction from websites. Octoparse\footnote{\url{https://www.octoparse.com/} } simplifies the data extraction process by enabling users to create a scraping workflow with clicks. It includes features like URL and string lists for targeted scraping and ready-to-use templates for popular sites like Amazon and Google. FactExtract  \cite{sarr2018factextract} is tailored for aggregating content from specific Senegalese news sources, boasting automatic language detection for ten languages, data cleaning, and analysis, all whilst avoiding data duplication. This tool, which utilises Python's Newspaper library, also features automated daily updates for the news content it monitors.
 \textit{ENoW} - News Data Extractor from the \textit{Web}\footnote{L Reips, M Musicante, G Vargas-Solar, ATR Pozo, C.S Hara, ENoW-Extrator de Dados de Notícias da Web,  Demonstration Anais Estendidos do XXXVIII Simpósio Brasileiro de Bancos de Dados, 2023, 78-83 
 } is a news scrapping system that explores online newspapers. 
\textit{ENoW} receives search strings as input and stores in a relational database data extracted from the news and their full content.  

\subsection{Data curation}
According to Garcov et al., \cite{garkov2023research}, research data curation is ``preparing research data and artefacts for sharing and long-term preservation''. Research repositories are the standard for publishing data collections to the research communities. Datasets at an early collection stage are generally not ready for analysis or preservation. Thus, extensive preprocessing, cleaning, transformation, and documentation actions are required to support usability, sharing, and preservation over time \cite{lafia2021leveraging}.  
Curated data collections have the potential to drive scientific progress \cite{zuiderwijk2020drives}, are relevant for reproducibility and improve the reliability of sciences \cite{vuorre2018curating}. However, data curation introduces challenges for supporting data-driven applications \cite{esteva2022synchronic} adopting quanti-qualitative methods. For example,  research challenges curating material across time, space and collaborators \cite{vuorre2018curating}. Quantitative and qualitative research methodologies apply ad-hoc data curation strategies that keep track of the data that describe the tools, techniques, hypothesis, and data harvesting criteria defined a priori by a scientific team.

Several software tools that apply statistical techniques and machine learning algorithms are available for qualitative researchers. Woods et al. \cite{woods2016researcher} argue that Computer-Assisted Qualitative Data Analysis Software (CAQDAS) is a well-known tool for qualitative research. These tools support qualitative techniques and methods for applying Qualitative Data Analysis (QDA). ATLAS.ti \cite{atlasti}, Dedoose \cite{dedoose}, MAXQDA \cite{maxqda}, NVivo \cite{nvivo}  implement the REFI-QDA standard, an interoperability exchange format. 
CAQDAS \cite{chen2018using}
researchers and practitioners can perform annotation, labelling, querying, audio and video transcription, pattern discovery, and report generation. Furthermore, CAQDAS tools allow the creation of field notes, thematic coding, search for connections, memos (thoughtful comments), contextual analysis, frequency analysis, word location and data analysis presentation in different reporting formats \cite{evers2018current}.  
The REFI-QDA (Rotterdam Exchange Format Initiative)\footnote{\url{https://www.qdasoftware.org}}
the standard allows the exchange of qualitative data to enable reuse in QDAS \cite{karcher2021data}. QDA software such as ATLAS.ti \cite{atlasti}, Dedoose \cite{dedoose}, MAXQDA \cite{maxqda}, NVivo \cite{nvivo}, QDAMiner \cite{qdaminer}, Quirkos \cite{quirkos} and Transana \cite{transana} adopt REFI-QDA standard.

We assume that data curation consists of identifying, systematizing,  managing, and versioning research data, considering versioning artefacts an essential component of tracking changes along the research project.

\subsection{Data labs}

Data science environments provide data labs like Kaggle\footnote{\url{kaggle.com}} and
Dryad\footnote{\url{https://datadryad.org/stash}} with stacks of services for (externalised) data storage, tagging and exploring tools. These environments
allow a collective sharing space of highly curated data collection maintenance tools. There are specialised repositories like DataOne\footnote{\url{https://www.dataone.org/about/}} and data repositories re3data\footnote{\url{https://www.re3data.org}}. 

DataONE (Data Observation Network for Earth) is a community-driven project that provides access to various environmental and ecological data across multiple member repositories. It is designed as an innovative framework aimed at facilitating research and enabling scientists and researchers to preserve, access, use, and increase the impact of their data. The platform provides robust data management tools, ensuring datasets' preservation and integrity. DataONE underscores data stewardship as a federated resource and supports scientific collaboration and reproducibility. It is invaluable for researchers seeking to address complex environmental challenges through shared data and knowledge. 

Re3data is a global registry of research data repositories that offers a comprehensive directory for researchers seeking to access, store, share, and manage their datasets. It represents a variety of academic disciplines and provides detailed information about each repository, such as access policies, standards, and contact details. re3data promotes data sharing, visibility, and reuse as a critical reference point for finding suitable repositories for data deposition. The platform enhances transparency in research data management. It supports open science by guiding users to trustworthy and reliable repositories, thereby facilitating the discovery of high-quality data across different scientific fields.

\subsection{Data lake, science lake and dataverse}

\paragraph{Data lakes} 
are expansive storage repositories that hold vast raw data in their native format until needed. Stein and Morrison \cite{giebler2019leveraging} emphasised their potential for scalability and flexibility in handling big data from various sources. In recent studies, Dixon in 2010\footnote{\url{https://jamesdixon.wordpress.com/2014/09/25/data-lakes-revisited/}} defined the term and its initial application in big data analytics. Quix et al. (2016) \cite{hai2021data} delved into the architectural considerations and challenges such as data governance and metadata management.

Science lakes, an offshoot of data lakes, are tailored specifically for the scientific community to address the need for interdisciplinary research, data management and complex analytics. Russom (2016) \cite{russom2016data} suggested that science lakes provide a more discipline-specific approach to data handling, enabling better metadata curation and domain-specific data models, which are crucial for reproducibility in scientific research.

A data lake is a vast storage system that houses extensive volumes of raw data in its original format. This versatile system accommodates a range of data types, including structured, semi-structured, and unstructured forms. Data lakes are essential in environments focused on big data analytics and are designed to manage data characterised by large volume, high velocity, and diverse variety from multiple sources. They are commonly utilised for advanced data processing activities such as machine learning and predictive analytics. Unlike traditional databases following the schema-on-write approach, data lakes follow the schema-on-read approach, providing flexibility in how data is formatted and used.

\paragraph{Dataverse.}

The concept of dataverse takes the notion of data lakes further by creating a networked space where data is stored, actively managed, and shared within the scientific community. 
A dataverse is a data repository platform for publishing, citing, and discovering datasets. It enables researchers to publish, cite, and discover datasets while providing metadata and tools to ensure others can understand and use data. Dataverses are often domain-specific and support the principles of open science, providing features such as data version control, digital object identifiers (DOIs) for citation, and tools for data analysis within the platform. They are community-driven and emphasize the accessibility and reusability of research data.

The most prominent example is the open-source Dataverse project developed by the Institute for Quantitative Social Science at Harvard University.
The Dataverse Project, initiated by King \cite{king2007introduction}, provides an open-source platform for sharing, preserving, citing, exploring, and analysing research data. It focuses on data citation and reproducibility, as discussed by Crosas \cite{crosas2015automating}, who highlighted the platform's role in fostering collaboration and open science.

Different academic institutions have built their dataverses for sharing and disseminating experimental scientific results, including the data collections they curate:
University of Arizona\footnote {\url{https://arizona.figshare.com}}, the
Different universities and academic institutions have promoted their dataverses like the University of Hamburg\footnote{\url{ https://www.fdm.uni-hamburg.de/en/fdm.html}}, the 
University of Michigan\footnote{ \url {https://www.icpsr.umich.edu/web/about/cms/2365}} and the
Grenoble Dataverse\footnote{\url{https://scienceouverte.couperin.org/cellule-data-grenoble-alpes/}}.

\paragraph{Summary.}
Together, these systems represent a shift toward more open, integrated, and efficient ecosystems for data management, offering novel solutions to the challenges posed by the vast amounts of data generated in modern research. They move away from traditional databases and toward more fluid, dynamic systems that can accommodate the ever-changing landscape of big data and scientific research.

A dataverse and a data lake are concepts related to data storage and management but serve different purposes and are designed with varying cases of use in mind.
While a dataverse is a scholarly platform aimed at curating, sharing, and preserving research data with rich metadata and community collaboration features, a data lake is a more generalised and scalable storage solution for raw data to support diverse data analytics and processing workflows.

\subsection{Data lakes and data verses in Life \& Earth sciences}

Dataverses in Life \& Earth sciences are specialised digital infrastructures designed to address specific data management needs for these scientific domains. They provide a structured yet flexible environment where datasets can be stored, accessed, shared, and analysed. These dataverses typically offer robust metadata standards and tools to ensure their data are well-described, making them discoverable and usable for various research purposes.

In Life Sciences, dataverses often focus on genomics, proteomics, clinical trials, and other biological data, integrating various sources of information to aid in complex analyses like phenotype-genotype correlations. For Earth Sciences, dataverses might concentrate on geospatial data, climate models, seismic activity records, and ecological data, supporting efforts to understand and model the Earth's dynamic systems.

These repositories support open science by promoting data sharing across disciplinary boundaries. This feature enables researchers to replicate studies and build upon existing work, which is fundamental for advancing knowledge. They also facilitate interdisciplinary collaboration, allowing experts from different fields to contribute to and draw from a collective data pool. For instance, a dataverse in these fields might include a combination of high-throughput experimental data, field observations, and simulation outputs.
The combination of openness and rigorous data management positions dataverses as critical resources in pursuing scientific discovery in Life \& Earth sciences.

In life and earth sciences, data lakes are pivotal for consolidating scientific data collected from various biodiversity studies and geological events like earthquakes. Once curated, processed, and analysed, this data contributes significantly to data-driven experiments underpinned by well-established protocols. The harvested data enriches the data lake and supports the creation of detailed, curated views for dissemination through dataverses. 

Our vision emphasises the importance of developing and maintaining data lakes with partially curated content in life and earth sciences, facilitating the continuous cycle of experimental data feeding back into the lake and subsequently sharing via dataverses.

\section{Maintaining and sharing earth and life sciences knowledge: challenges}\label{sec:approach}

 Various data on life and earth sciences have been acquired from different sources \cite{da2022using}.
Integrated access to data collections and their curated versions can facilitate their maintenance, analysis and experimentation.
It can also demonstrate knowledge of the discipline with 
its vocabulary, concepts and relationships in a synthetic way.

 Curation, maintenance and exploration of data collections in the data lake calls for proposing
techniques for exploring data collections that can be explored and enriched while producing
new data and analytical results. Data curation also means keeping track of the 
type of experiments carried out on the data, their results and the conditions under which they were
carried out. Maintaining a catalogue of data-related questions and experiments can promote open science, share data and knowledge, and share the data and knowledge the scientific community has gained from it \cite{adorjan2022towards}. This information
should also be stored in the data lake.

\paragraph{Challenge 1: How to structure and organise life and earth sciences metadata?}
Metadata modelling is a way of structuring and organising earthquakes and biodiversity. The metadata
 model must make the content of a data lake findable, accessible, interoperable and reusable
(FAIR principles \cite{wilkinson2016fair}). Metadata can represent the data's structural, semantic and contextual aspects (provenance, conditions and assumptions under which the analytical results are obtained, i.e., the metadata driving the analysis). Most proposed models are based on logic or structured by graphs \cite{scholly2021goldmedal,diouan2022metadonnees} that can be
specialised in seismic geophysical data and biodiversity. Besides, associating metadata can be achieved by considering quantitative and qualitative perspectives through data curation. Combining quantitative and qualitative approaches allows for a meta-model of the content used and produced in experiments and the conditions in which the content is produced, chosen, validated and considered representative knowledge for the domain of study.


\paragraph{Challenge 2: How to integrate data in the data lake?}
Since the experiments require several data collections, integrating the data into the data lake must be part of a pipeline that includes data discovery, exploration, selection and integration. This process should be designed
based on the requirements of life and earth science experiments \cite{da2022using}. The heterogeneity of the data (text, signals, multimedia, proprietary formats from
seismographs), the speed of the data often produced in the form of streams in the case of seismic sensors
in addition to the volume are aspects that require original contributions in the design, maintenance and exploration of the data lake.

\paragraph{Challenge 3: How to integrate data in the data lake considering scientists' needs?}
The researcher's intervention, defined as a researcher-in-the-loop (RITL) \cite{van2020researcher}, is a crucial aspect of human intervention to assess content concerning (i) the conditions in which it is produced and (ii) to make decisions about the new tasks to perform and the way a research project will move forward. RITL is a case of Human-in-the-loop (HITL), where the primary output of the process is a selection of the data, not a trained machine-learning model. HITL is crucial for handling supervision, exception control, optimisation, and maintenance \cite{rahwan2018society,mosqueira2023human}. Under a RITL approach, a human sees all data points in the relevant selection at the end of the process. Using RITL requires a systematic solid way of working\footnote{\url{https://hai.stanford.edu/news/humans-loop-design-interactive-ai-systems}}.
This characteristic is critical for designing content curation for quantitative and qualitative research methods.

Scientific content should be extracted and computed, including data, analytics tasks (manual and AI models), and associated metadata. This curated content allows the produced knowledge to be reusable and analytics results to be reproducible \cite{leipzig2021role}, thereby adhering to the FAIR principles \cite{barcelos2022fair}.

\section{Towards a curation approach for building a Life \& Earth sciences data lake}\label{sec:platform}
\begin{figure*}[t]
   \centering
\includegraphics[width=0.95\linewidth]{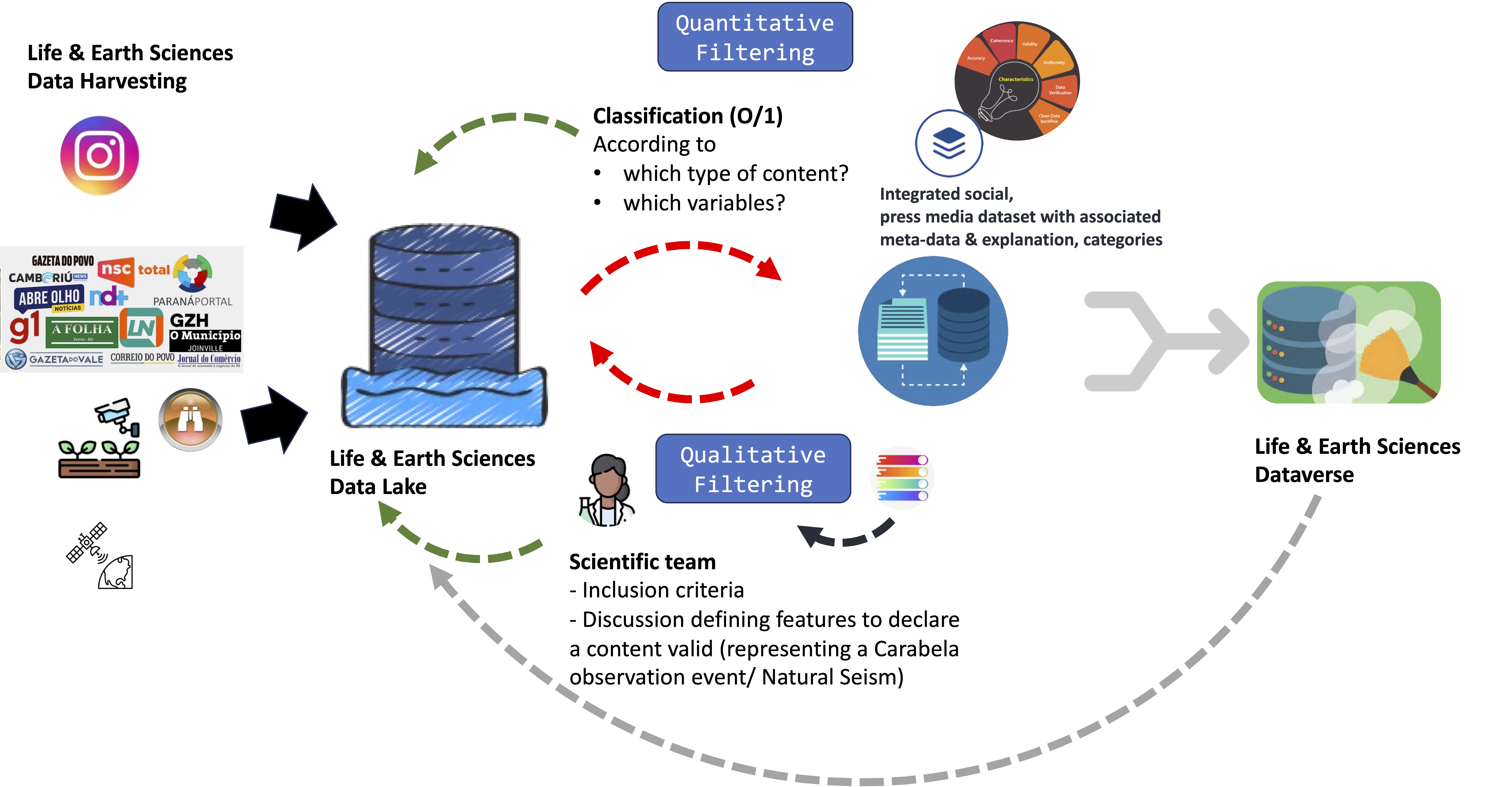}
 \caption{General overview of the curation approach for building, maintaining and exploiting a data lake.}
   \label{fig:architecture}
 \end{figure*}

Figure \ref{fig:architecture} illustrates the principle of our vision concerning the way a life and earth sciences data lake can be built, maintained and exploited. Our approach is based on the quantitative and qualitative curation of data harvested digitally and {\em in situ} (left-hand side of the figure). Heterogeneous raw data is gathered and stored in the data lake. Then, algorithms (statistical and Artificial Intelligence) and researchers can process, filter and classify data. This filtering process produces and stores meta-data in the data lake.
Data exploration and integration (cleaning and engineering) processes can be performed on data samples from the data lake. They can be used for experimental purposes to produce content associated with the data stored in the data lake. Clean and curated data associated with meta-data representing the quantitative and qualitative perspective of the experiments can then be shared in a data verse (right-hand side of the figure).

\paragraph{ Harvested data, models and knowledge integration.} 
Various life and earth sciences data have been harvested from different sources. Since they are heterogeneous and produced at different paces (continuous and in batch), our approach proposes an integration approach based on a pivot meta-representation. The principle is to present a general meta-model of their content and process them for extracting technical, structural and semantic meta-data. This abstract representation provides integrated access to data collections and curated versions under a global knowledge graph and can promote their maintenance, analysis, and experimentation. It can also show the knowledge of the discipline with its vocabulary, concepts, and relations in a synthetic manner. The data lake can be pivotal in collecting, processing, and exporting raw data in a curated view.

\paragraph{ Curation, maintenance, and exploration of data collections for bringing data value from in situ observations and experiments.} 
Since data acts as a backbone in modelling phenomena for understanding their behaviour, it is critical to developing good collection and maintenance: which are available data collections? Are they complete? Which is their provenance? In which conditions were they collected? Have they been processed? In which cases have they been used, and what are the associated results?
We propose techniques to explore data collections using graphs that can be explored and enriched while new data and analytics results are produced. Data curation also means keeping track of the type of experiments run on data, their results, and the conditions in which they were performed. Maintaining a catalogue of data-related questions and experiments can promote open science and share data and the knowledge that the scientific community has derived from it. 

\paragraph{ Modelling and simulating experiments to answer questions in life and earth sciences. } 
Answering research questions through data-driven experiments implies:
\begin{itemize}
    \item Designing ad hoc experiment artefact models and programming languages for enabling friendly, context-aware, and declarative construction of experiments in life and earth sciences.

    \item Collecting execution of experiments data (raw input data, prepared datasets,  experiments’ tasks calibration and associated results).

\end{itemize}
	
\paragraph{Pilot experiments.}
The data lake will be tested in real scenarios through collaboration with domain experts in seismology and biodiversity studies in Brazil. 
The entry point will be two pilot experiments, namely:
\begin{enumerate}
    \item  the classification process of seismic signals collected by stations through different observations to detect "natural" and human-made earthquakes in the northern human-made earthquakes in the northern region of Brazil;

    \item  the classification of \emph{in situ} observations of the "carabela portuguesa"\footnote{The Portuguese caravel (Physalia physalis) is a monotypic colonial species of siphonophore hydrozoan of the family Physaliidae. It is commonly found in the open ocean in all warm waters of the world, especially in the tropical and subtropical regions of the Pacific and Indian Oceans, as well as in the Atlantic Gulf Stream. Its sting is dangerous and very painful \url{https://es.wikipedia.org/wiki/Physalia_physalis}.} and modelling its behaviour on the Brazilian coast.
    
\end{enumerate}
 
In both cases, it is necessary to (i) apply statistical methods to investigate and unveil new patterns in seisms and biodiversity data, answering open problems or leading to new research questions; (ii) build predictive models to better describe or approximate phenomena, increasing the knowledge about our planet. The conditions in which statistics and prediction are performed, results, observations, interpretation and validation of the results are data to be integrated into the data lake.

\paragraph{Discussion.}
The originality of the work is to address the construction of a data lake that  includes:

\begin{enumerate}

    \item  Raw collected data representing life and earth sciences phenomena (streams, batch, multimedia, proprietary).

    \item  Data produced along data-driven experiments adopting data science techniques including artificial intelligence algorithms (ML-driven data lakes).

    \item  Contextual data describing the conditions in which data are collected, and experiments are designed and enacted. The data lake will provide data curation modules for extracting metadata according to a well-adapted model and modules exploring data and using them for designing new experimentations, thereby adopting an open science perspective.
    
\end{enumerate}

\section{Conclusions and future work}\label{sec:conclusion}
Our vision is that it is necessary to address fundamental research topics at the centre of Data Science, Big Data management and analytics for solving data-driven problems in life and earth sciences.

The contribution is the design and exploration techniques of a data lake with a well-adapted model for metadata about life and earth sciences experiments consuming and producing quantitative and qualitative data. An important work will be to define exploration operators and pipelines to exploit the content for further maintaining and developing new  life and earth sciences experiments.

\section{Acknowledgements}

The work reported in this paper is done in the context of the LETITIA\footnote{\url{http://vargas-solar.com/letitia/}} project, funded by the \emph{Fédération Informatique de Lyon}\footnote{\url{https://fil.cnrs.fr}}. 

\bibliographystyle{ACM-Reference-Format}
\bibliography{sample-ceur}

\end{document}